
\input harvmac
\def\bd {{\bar \del}}\def \ra {\rightarrow}
\def \vp {\varphi}
\def \pw {`plane wave'\ }

\def \eq#1 {\eqno{(#1)}}

\def \a {\alpha}

\def \b {\beta}
\def \k {\kappa}

\def \p {\phi}

\def \d {\delta}
\def \l {\lambda}
\def \m {\mu}
\def \g {\gamma}
\def \n {\nu}

\def \fourth {{1\over 4}}

\def \e#1 {{\rm e}^{#1}}
\def \const {{\rm const }}

\def \vp {\varphi}

\def \dg  {{\dot g}}
\def \ddg  {{ \ddot g}}

\def \ha { { 1\over 2 }}

\def \ov {\over}

\def \sm  { sigma model\ }

\def\np {  Nucl. Phys. }
\def \pl { Phys. Lett. }
\def \mpl { Mod. Phys. Lett. }
\def \prl { Phys. Rev. Lett. }
\def \pr  { Phys. Rev. }
\baselineskip8pt
\Title{\vbox
{\baselineskip8pt{\hbox{CERN-TH.7069/93}}{\hbox{PRA-HEP 93/15}  }
{\hbox{hep-th/9311012}} }}  {\vbox{\centerline
{ Duality invariant class of exact string backgrounds }
}}
\vskip -10 true pt
\centerline  {   C. Klim\v c\'\i k\footnote{$^{*}$}{\baselineskip5pt
e-mail:  presov@cspuni12.bitnet} }
\vskip 2pt
\centerline {\it  Theory Division, Nuclear Centre, Charles University,  }
\centerline {\it   180 00 Prague 8, Czech Republic}
\vskip 3pt
\centerline {and}
\vskip 5pt
\centerline{   A.A. Tseytlin\footnote{$^{**}$}{\baselineskip5pt
On leave  from Lebedev  Physics
Institute, Moscow, Russia.
e-mail: tseytlin@surya3.cern.ch  and  tseytlin@ic.ac.uk} }
\vskip 2pt
\centerline {\it Theory Division, CERN}
\centerline {\it
CH-1211 Geneva 23, Switzerland}
\vskip 2pt
\centerline  {\it and}
\vskip 2pt
\centerline {\it   Theory Group, Blackett Laboratory, Imperial College}
\centerline {\it London SW7 2BZ, U.K.}
\vskip 14pt
\vskip 3pt
\baselineskip7pt
\noindent
We consider a class of $2+D$ - dimensional string backgrounds with a
target space metric  having a covariantly constant null Killing vector
and flat `transverse' part. The corresponding sigma models are invariant
under $D$ abelian isometries and are transformed by  $O(D,D)$ duality
into models belonging to the same class. The leading-order solutions
of the conformal invariance equations (metric, antisymmetric tensor and
dilaton), as well as the action of  $O(D,D)$ duality transformations on
them, are exact, i.e. are not modified by $\a'$-corrections. This makes
a discussion of different space-time representations of the same string
solution (related by $O(D,D|Z)$ duality subgroup) rather  explicit.
We show that the $O(D,D)$ duality may  connect  curved  $2+D$-dimensional
backgrounds  with solutions  having  flat metric but, in general,
non-trivial antisymmetric tensor and dilaton. We discuss several
particular  examples including the $2+D=4$ - dimensional background
that was recently  interpreted in terms of a WZW model.

\vskip 20pt
\vskip 2pt
\noindent
{CERN-TH.7069/93}

\noindent
{November  1993}
\Date { }

\noblackbox
\overfullrule=0pt
\baselineskip 20pt plus 2pt minus 2pt

\def\np {  Nucl. Phys. }
\def \pl { Phys. Lett. }
\def \mpl { Mod. Phys. Lett. }
\def \prl { Phys. Rev. Lett. }
\def \pr  { Phys. Rev. }

\lref \berg {E.A. Bergshoeff, R. Kallosh and T. Ortin, \pr D47(1993)5444. }

 \lref \kumar {A. Kumar,
\pl B293(1992)49; D. Gershon, preprint TAUP-2005-92.}
\lref  \hussen {  S. Hussan and A. Sen,  \np B405(1993)143. }
\lref \kiri {E. Kiritsis, \np B405(1993)109. }
\lref \alv { E. Alvarez, L. Alvarez-Gaum\'e, J. Barb\'on and Y. Lozano,
preprint CERN-TH.6991/93.}
\lref \sfts { K. Sfetsos, to appear. }
\lref \nap { M. Henningson and C. Nappi,  \pr D48(1993)861.  }
\lref \Ve    { K.M. Meissner and G. Veneziano, \pl B267(1991)33;
A. Sen,  \pl  B271(1991)295.}

\lref \givkir {A. Giveon and E. Kiritsis,  preprint CERN-TH.6816/93,
RIP-149-93. }

\lref  \rocver { M. Ro\v cek and E. Verlinde, \np B373(1992)630.}

\lref \brink{ H.W. Brinkmann, Math. Ann. 94(1925)119.}
\lref \guv {R. Guven, Phys. Lett. B191(1987)275.}

\lref \amkl { D. Amati and C. Klim\v c\'\i k, \pl B219(1989)443.}

\lref \hor { G. Horowitz and A.R. Steif, Phys.Rev.Lett. 64(1990)260 ;
Phys.Rev. D42(1990)1950.}

\lref \horr {G. Horowitz, in: {\it Proceedings
of  Strings '90},
College Station, Texas, March 1990 (World Scientific,1991).}

\lref \rudd { R.E. Rudd, \np B352(1991)489 .}

\lref \duval { C.Duval, G.W. Gibbons and P.A. Horv\'athy, \pr D43(1991)3907 ;
C.Duval, G.W. Gibbons, P.A. Horv\'athy and M.J. Perry, unpublished (1991);
C. Duval, Z. Horvath and P.A. Horv\'athy,  \pl B313(1993)10; Marseille
preprints
 (1993). }

\lref \tsnul { A.A. Tseytlin, \pl B288(1992)279; \pr D47(1993)3421.}
\lref \tsnull { A.A. Tseytlin, \np B390(1993)153.}

\lref \dunu { G. Horowitz and A.A. Steif, \pl B250(1990)49;
 E. Smith and J. Polchinski, \pl B263(1991)59;
J. Horne, G. Horowitz and A. Steif, \prl 68(1992)568;
G. Horowitz, in; {\it  Proc. of the 1992 Trieste Spring School on String theory
and Quantum Gravity},
preprint UCSBTH-92-32, hep-th/9210119;
G. Horowitz and D.L. Welch, \prl 71(1993)328.
}

 \lref \busc { T.H. Buscher, \pl B194(1987)59 ; \pl B201(1988)466.}
\lref \pan  { J. Panvel, \pl B284(1992)50. }

\lref \mye {  R. Myers, \pl B199(1987)371;
    I. Antoniadis, C. Bachas, J. Ellis, D. Nanopoulos,
\pl B211(1988)393;
 \np B328(1989)115. }
\lref \givpas { A. Giveon and A. Pasquinucci, \pl B294(1992)162. }
\lref \GK {A. Giveon and E. Kiritsis,  preprint CERN-TH.6816/93,
RIP-149-93.}
\lref \givroc {A. Giveon and M. Ro\v{c}ek, Nucl. Phys. B380(1992)128.}
\lref \giv {  A. Giveon, Mod. Phys. Lett. A6(1991)2843;  R. Dijkgraaf, H.
Verlinde and
E. Verlinde, \np B371(1992)269; I. Bars, preprint USC-91-HEP-B3;
 E. Kiritsis, \mpl A6(1991)2871. }

\lref \GRV {A. Giveon, E. Rabinovici and G. Veneziano, Nucl. Phys.
B322(1989)167;
A. Shapere and F. Wilczek, \np B320(1989)669.}
\lref \GMR {A. Giveon, N. Malkin and E. Rabinovici, Phys. Lett. B238(1990)57.}
\lref \Ve    { K.M. Meissner and G. Veneziano, \pl B267(1991)33;
M. Gasperini, J. Maharana and G. Veneziano, \pl B272(1991)277; \pl
B296(1992)51.}
\lref \sen {A. Sen,  \pl  B271(1991)295; \pl B274(1991)34.}

\lref \VV    {  G. Veneziano, \pl B265(1991)287.}

\lref  \horne { J.H. Horne and G.T. Horowitz, \np B368(1992)444.}

\lref \koki { K. Kounnas and  E. Kiritsis, preprint CERN-TH.7059/93;
hep-th/9310202. }
\lref \napwi { C. Nappi and E. Witten, Princeton preprint IAS-HEP-93/61;
hep-th/9310112. }

\lref \tsmpl {A.A. Tseytlin, \mpl A6(1991)1721.}

\lref \tsbh {A.A. Tseytlin,  preprint CERN-TH.6970/93; hep-th/9308042.}

\lref \nsw {  K.S. Narain, M.H. Sarmadi and E. Witten, \np B279(1987)369. }

\lref \STT {K. Sfetsos and A.A. Tseytlin, preprint CERN-TH.6969/93,
hep-th/9310159.}
\lref \givkir {A. Giveon and E. Kiritsis,  preprint CERN-TH.6816/93;
hep-th/9303016. }

\lref \cec { S. Cecotti, S. Ferrara and L. Girardello, \np B308(1988)436. }

\lref \tsdu { A.A. Tseytlin, \np B350(1991)395.  }

\lref \wit {E. Witten, Commun. Math. Phys. 92(1984)455 ;
E. Braaten, T.L. Curtright and C.K. Zachos, \np B260(1985)630.}



1.  \ The aim of the present  paper  is to   study the action of the  duality
transformations
 on a  simple class of  exact  string  solutions  which have a covariantly
constant null Killing
vector. Such (`plane wave' type) solutions of Einstein equations  are well
known \brink.
 Some particular examples of such spaces were found to be  solutions of the
string effective equations
to all orders of $\a'$ perturbation
theory   \guv\amkl\hor\horr\rudd\duval\berg. In the case when the `transverse'
space is
not  described by a conformal theory (e.g. is not flat)  one finds more
general solutions
with  the `transverse' couplings  satisfying  the  first order
renormalisation group - type equations (with  the light-cone coordinate playing
the
role of the RG time) \tsnul\tsnull.
The advantage of the \pw backgrounds is their simplicity and  controllable (or
vanishing)
quantum $\a'$-corrections. This suggests that duality transformations acting on
such backgrounds
should also  take  an explicit form.\foot{Simple one-isometry duality
transformations of the \pw
backgrounds were previously discussed in \dunu.}

As it is well known, string  solutions with $D$  abelian isometries are
transformed into each other
by  the  $O(D,D|R) $ duality rotations (see, e.g.,
\GRV\GMR\busc\tsmpl\VV\Ve\sen\rocver\givroc\alv).  The origin of this symmetry
is  in the $2d$
duality  on the string world sheet that relates `left' and `right' parts of the
isometric string
coordinates  leading to the $O(D)\times O(D)$ symmetry in the correlators of
vertex operators and
thus (extended off shell)  in the string field theory \sen\ or string effective
  action.
 Combining the $O(D)\times O(D)$ rotations  with the obvious  $GL(D,R)$
coordinate
transformations and  constant shifts of the antisymmetric tensor  one finds
the $O(D,D|R) $ duality
group.
 This   symmetry  becomes manifest  in  the phase-space approach  \GRV\Ve\ or
closely related
(non-manifestly Lorentz invariant)  Lagrangian  approach  \tsdu.  From the \sm
path integral point
of view,  one is able to make the formal  $2d$ duality transformation (via
introducing Lagrange
multipliers or gauging) that transforms one \sm into its $O(D,D|R) $ rotation
\busc\rocver\givroc.
Under special conditions (compactness of orbits of isometries) the
transformations from the
$O(D,D|Z) $  subgroup relate the backgrounds that correspond to the same
conformal field theory
 \rocver\givroc.

In general, the leading-order  duality preserves only the one-loop  conformal
invariance equations,
 i.e. transforms
the leading-order string  solutions into the  leading-order ones (or exact
solutions into
leading-order ones). In fact, the leading-order duality transformation rules
are known to be modified
by $\a'$-corrections \tsmpl\pan. However,  such $\a'$-corrections may be absent
 for certain
backgrounds. Then the leading-order duality transformations would relate exact
string vacua. A recent
example   is provided by  the duality in the $SL(2,R)/R\ $    $ 2d$ `black
hole'
background \giv\tsmpl\ which remains semiclassical in a specific scheme \tsbh.
Similar result  is
true  for the duality in the $[SL(2,R)\times R]/R$  gauged WZW model ($3d$
`black string' \horne)
which also retains its semiclassical form in a special scheme  \givkir\STT.

2. \  Below  we shall present a  simpler
and more explicit  example  of  a  situation when the leading-order duality
transformations are
actually exact.
Consider a \sm with
coordinates $x^i$  describing a flat  Euclidean signature (or toroidal \nsw)
 background  parametrised by constants  $g_{ij}, \ b_{ij}, \ \p\  (i,j=1,...,D)
$.
Coupling  it to  an additional \sm  with coordinates $t^a$  (i.e.
making $g,b,\p$  dependent on $t^a$) one gets a   class of models  with $D$
abelian isometries
which   is invariant under $O(D,D|R) $ duality.\foot{
 In general, one can
also introduce `non-diagonal' terms  like $A_{ai} \del t^a\bd x^i$ (see e.g.
\givroc) which we shall
ignore.}
One obvious  possibility is to consider a class of cosmological backgrounds by
taking $t^a$ to be just
one time coordinate.  The resulting  leading-order term in the string effective
action and the
corresponding field equations for $g(t),\ b(t), \ \p(t)$ are explicitly
invariant under the duality
\Ve. However, both the solutions and their duality rotations in this case are
modified  by
$\a'$-corrections and their exact form is not known explicitly.

Instead,   we shall take  $t^a$ to be the pair  of  light-cone coordinates
$(u,v)$
and impose the null Killing symmetry by letting
 the background fields to depend only on $u$, i.e.
$$I(u,v,x) =   {1 \over \pi \a' } \int d^2 z  \{  -2 \del u \bd v \ + \
 ({ g}_{ij } + b_{ij})(u) \ \del x^i\bd x^j\}   +  {1 \over 4 \pi  } \int d^2 z
 \sqrt \g R^{(2)}
\p(u) \ .  \eq{1} $$
The only non-zero components of the connection, curvature and $H_{\m\n\k}=
(\del_\m b_{\n\k} +
cycle) \ (x^\m= u,v,x^i)$ are
$$ { \Gamma}^v_{ij}=\ha \dg_{ij}
\ , \ \ \ { \Gamma}^i_{ju}=\ha g^{ik}\dg_{kj}   \ \ , \ \
\ {\dot F }\equiv{d  F  \over d u}\ \ ,  $$
 $$
{ R}_{iuju}= -{1\over 2}  \ddg_{ij} + {1\over 4}  g^{mn}\dg_{im}\dg_{nj} \  , \
 \ \ \
  H_{uij} = {\dot b}_{ij} \ \ , \eqno (2) $$
so it is easy to show  that the one-loop conformal invariance equations
$$\ {\bar \b}^G_{\m\n} = R_{\m\n} - \fourth H_{\m\k\l} H_{\n}^{\ \k\l} + 2 D_\m
D_\n \p =0\ , \ \ \
{\bar \b}^B_{\m\n} = - \ha D^\l H_{\l\m\n} + \del^\l \p H_{\l\m\n} =0 \ , $$
reduce just to {\it one}
equation (${\bar \b}^G_{uu}=0$)\foot{
  The
most general $D+2$ dimensional Minkowski signature metric
 admitting a covariantly constant null Killing
vector can
be represented as  $ds^2 =  G_{\m\n}dx^\m dx^\n =  -2dudv
+  g_{ij} (u,x) dx^i dx^j \  . $
When the transverse metric is flat one can make
 a coordinate transformation to put the full metric  into the form:
$ds^2 =   -2dudv +   dx^i dx_i + 2 A_i(u,x) dx^i du + K(u,x) du^2  \ . $
As explained  in \tsnull\  exact solutions of the string equations with
$g_{ij}=\d_{ij} , \ A_i= - \ha F_{ij} (u) x^j ,\  K=k_{ij} x^i x^j +k_0, $
considered in
\amkl\hor\horr\ are equivalent to the solutions  with
$g_{ij} = g_{ij}(u)$ in \rudd\duval.
Equivalent representations for $B_{\m\n}$  are  $B_{iu}= - \ha {\dot b}_{ij}x^j
, \ B_{ij} =0 $
and  $B_{iu}= 0 , \ B_{ij} = b_{ij} (u) $.
Note that in  our model (1)   $B_{\m v}=0$ since if  $B_{\m v}= B_{\m v}
(u)\not=0$
there are non-vanishing higher order $\a'$-corrections to the string field
equations.
}
$$ -{1\over 2} g^{ij}  \ddg_{ij} + \fourth  g^{ij} g^{mn}\dg_{im}\dg_{jn}
 - \fourth g^{ij} g^{mn}{\dot b}_{im}{\dot b}_{jn}
 + 2 \ddot \p =0\ .
\eq{3} $$
Moreover, there are no $\a'$-corrections to the ${\bar \b}^{G}  ={\bar
\b}^{B}=0$ equations on this
background:
  all
higher-order  contractions  of relevant tensors vanish   \horr\tsnull.  This is
also obvious
from the path integral point of view \amkl : the field $v$ in (1) plays the
role of a Lagrange
multiplier  that `freezes' out  fluctuations of $u$ unless there are sources
corresponding to the
$u$-direction. Once the one-loop condition of conformal invariance in the
$uu$-direction  (3) is
satisfied, it is easy to argue  from the structure of (1) that no higher-loop
divergences can appear.
Since all the scalar invariants vanish, there are no contributions to the
central charge, i.e.
$c= 2 + D$.\foot{In contrast to the case of non-conformal `transverse' theory
\tsnul\tsnull\
 here  one  cannot include the linear  $pv$-term  in $\p$ since this  leads  to
the conditions $p{\dot g}_{ij} =0, \  p{\dot b}_{ij} =0$.}

The solutions of (3) thus represent {\it exact}  string backgrounds.
 Since we have just one equation for $D^2 +1$ unknown functions $g_{ij} (u),\
b_{ij}(u), \ \p(u) $
there are many particular  solutions. For example, we  can set $b_{ij} =
\const$ or $\p=const$
or   $g_{ij} = \d_{ij}$.  The latter case looks  the simplest  since
it corresponds to the {\it flat}  $2+D$ dimensional space  (and in this sense
is analogous to
the linear dilaton background \mye) with
 the conformal invariance (3) being  maintained by  the balance of
contributions
of non-constant
antisymmetric tensor $and$ dilaton fields. In contrast to the case of the
linear dilaton vacuum
here the propagation of the classical string is non-trivial  being effected by
the $b_{ij}$
background. As for the propagation of the quantum string modes,
it  is easy to see that the propagation of
the tachyon field in this background  remains the same as in  empty space: the
dilaton dependence
in  $$  - \ha    D^2 T +  \del^\m \p \del_\m T - {2\ov \a'} T = -\ha \del^2_i T
- \del_u\del_v T  +
\del_u\p\del_v T  - {2\ov \a'}  T $$
 can be completely eliminated by $T\ra T' = {\rm e}^{\p(u)} T$. Similar
conclusion holds  for   the propagators of other modes except the graviton one
(the graviton
propagator is  changed from the flat one by $O({\dot b}_{ij}^2)$-dependent
terms).  The string
interaction vertices  depend of course  on $\p(u)$ (and some also on  ${\dot
b}_{ij}$) so  that the
scattering of strings on this background should  be quite non-trivial.

Another special case is $\p=\const$. Then eq.(3) is satisfied due to the
cancellation between the
Ricci tensor and the  antisymmetric tensor field strength contributions as for
parallelisable
spaces or WZW models \wit.  In fact,  a particular model of this type (with
$D=2$) was
recently interpreted \napwi\ as a WZW theory based on a non-semisimple group
(a central extension of the 2-dimensional Euclidean group).
 The
corresponding action   can be put into the form
$$I(u,v,x) =   {1 \over \pi \a' } \int d^2 z  \bigl(  -2 \del u \bd v +
 \del x_1\bd x_1 + \del x_2  \bd x_2  + 2  \cos u \del x_1 \bd x_2 \bigr) +  {1
\over 4 \pi  }
\int d^2 z  \sqrt \g R^{(2)}
 \p_0 \ .
\eq{4} $$
The  structure of this action  is very simple: if one formally replaces $-2\del
u \bd v$ - term
by $\del u \bd u $  then (4) becomes  the action of the $SU(2)$ WZW model. One
can also   make the
complex rotation $u'=iu, \ v'=-iv$ replacing  $\cos u$ by $\cosh u$  and thus
going from  $SU(2)$
 to $SL(2,R)$.  This observation implies that the $O(2,2)$  duality
transformations  of (4) with
respect to $x_1,\ x_2$ look   the  same (with $\del u \bd u \ra -2\del u \bd
v$)  as the
  duality rotations of the  $SU(2)$ WZW model  (discussed in
\rocver\kumar\hussen\givroc\kiri).
 For example,
performing  the duality transformation in  the $(x_1+x_2)$-direction we get
$$I' =   {1 \over \pi \a' } \int d^2 z  \bigr[  -2 \del u \bd v +
  {\rm tg}^2  ({u\over 2})  \del y_1\bd y_1 + \del y_2  \bd y_2 \bigl]
 +  {1 \over 4 \pi  } \int d^2 z  \sqrt \g R^{(2)}[\p_0 - \ha \ln ( 1+ \cos u)
]  \ .
\eq{4'} $$
Another example is a partucular $O(2,2)$ duality transformation \givroc\ that
leads from
from the  `charged black hole' ($[SL(2,R)\times R]/R $ gauged WZW) model of
\horne
$$ I'' =   {1 \over \pi \a' } \int d^2 z  \{  -2 \del u \bd v
 + {(\l-1)\sin^2
{u\over 2} \over \cos^2 {u\over 2} - \l }  \del y_1 \bd y_1  + {\l \ \cos^2
{u\over 2} \over  \cos^2 {u\over 2} - \l    }  \del y_2 \bd y_2 \   $$
 $$ -{\l \ \sin^2
{u\over 2} \over  \cos^2 {u\over 2} - \l   }( \del y_1 \bd y_2 -
\del y_2 \bd y_1 )  \}
 + {1 \over 4 \pi  } \int d^2 z  \sqrt \g R^{(2)} [\p_0 - \ha \ln ( 1-2\l +
\cos
u)]   \ ,  \eq{4''} $$
where $\l$ is a free parameter.  The models (4$'$),(4$''$) belong to the same
class (1). Since the
standard leading-order duality
rotations  we have applied  preserve the one-loop  string  equations
(i.e. preserve (3), see below)  the  backgrounds in  (4$'$),(4$''$)   also
represent exact
string solutions.

3. \ Let us now discuss  systematically the  action of the $O(D,D|R)$ duality
on the above class (1)
of exact string solutions. First, let us  demonstrate explicitly that eq.(3) is
invariant under
$O(D,D|R)$ duality. The argument is essentially the same as in the
time-dependent case of \Ve\ (but
even simpler).\foot{ It is possible to check  the invariance of the
string effective action as in \Ve\  by introducing  the Lagrange--multiplier
field
$G^{uu}$ (not transforming  under the duality) the variation over  which gives
(3).
Then the  string effective action takes the form:
$$ \int du \e{-2\vp }   \{-{ 2\ov 3}(D-24)  +
\a' G^{uu}[{1\ov 8} \Tr \bigl( {\dot M }\eta {\dot M }\eta ) + 2
\ddot \vp ] \} \ .  $$.  }
 Introducing the basic $2D\times 2D$  matrix $M(u) $  \GRV\  built out of
$g_{ij}(u)
$ and $b_{ij}(u)$, the  constant matrix $\eta$  and the duality-invariant
dilaton $\vp(u)$
\VV\tsmpl\  $$M =\pmatrix{ g\inv &  -g\inv b \cr   bg\inv & g- b g\inv b  \cr}\
,
\ \ \ \ \eta = \pmatrix{ 0 &  I \cr   I&  0   \cr}\ ,
\eq{5} $$
$$      \vp = \p - \fourth
\ln \det g \ ,  \ \ \ \ \ \  \e{-2\vp }  =   \e{-2\p }  \sqrt g\ \ ,  \eq{6} $$
one can represent (3) in the form
$$ {1\ov 8} \Tr \bigl( {\dot M }\eta {\dot M }\eta ) + 2 \ddot \vp =0 \ ,
\eq{7} $$
which is manifestly invariant under the $O(D,D|R)$ transformations
(with a constant  parameter matrix $\Lambda$)
$$ M' = \Lambda M \Lambda^T\ , \ \ \ \vp'= \vp \ , \ \ \ \  \Lambda^T \eta
\Lambda = \eta\ . \eq{8}
$$
The two terms in (7) are particular invariants  of the  duality group.
The class of backgrounds corresponding to (1) splits into orbits under the
action  (8) of $O(D,D|R)$.
For example, let us consider the orbit that contains the flat background
$g_{ij}=\d_{ij}$.
It is clear that  (some)  duality transformations will rotate this flat
background into  curved ones.
This is similar to the well-known example of the duality transformation  acting
on flat space in
flat coordinates \rocver\Ve: $ ds^2= dr^2 + r^2 d\theta^2 \ra ds^2= dr^2 +
r^{-2 }d\theta^2 $.
However, in our case both the original and the duality transformed backgrounds
are known explicitly
since they  are not modified by $\a'$-corrections.
For example, consider the following $D+2=4$  background
$$g_{ij}=\pmatrix{1&f\cr f& 1+f^2\cr}\ , \ \ \ b_{ij} =0 \ , \ \ \  f=f(u) \ ,
\ \ \p = \p(u) \ ,
\eq{9} $$
where $f$ and $\p$ are related by (3), i.e.
$R_{uu}=-{1\over 2}\dot{f}^2 = - 2 \ddot \p .$  After the    duality  rotation
 in
the  $x_1$-direction  we get
$$ g'_{ij} = \d_{ij} \ , \ \ \ \ \ b'_{ij} = - f\epsilon_{ij} \  , \ \ \ \
\p'= \p  \ , \eq{10} $$
i.e. we have  transformed  a curved  ($R_{iuju}\not=0$) four-dimensional
background into  the
 one with a  flat  metric (but non-trivial antisymmetric tensor and dilaton).

If we assume that the spatial directions $x^i$ form a torus then the  duality
subgroup $O(D,D|z)$
will generate backgrounds that all correspond to the same conformal field
theory \rocver\givroc.
In particular,   (9) and (10)   thus  give different space-time
representations of the {\it same }  exact string  solution.
Another  example is provided by  (4) and (4$'$) and (4$''$).

4. \  In addition to the  $x^i$-isometries,  the action (1) is also invariant
under the constant
shifts of $v$ corresponding to the null Killing vector.
 This is a novel situation absent in the case of the
`cosmological' backgrounds of \Ve.
 One is thus tempted to consider more general $O(D+1,D+1)$ rotations of the
models (1).
Performing the duality `inversion' in the null  direction  leads to a
singularity.
One could  hope that  inclusion of  other non-null isometry directions may
`regularize' this singularity. Formally, the string equations should still be
invariant under
such duality rotations.  In  general,  $O(D+1,D+1)$ transformations
 move us out of our  class of exact  null  backgrounds (1). However, a subset
of transformations
which  contains  $O(D,D)$ transformations combined with `triangular'
$GL(D+1,R)$  coordinate
transformations $ v'= av + c_i x^i, \ {x'}^i= d^i_j x^j $  still preserves the
class (1).

 Let us now show that first making a coordinate transformation and then  such
an  $O(3,3)$
duality rotation   including the $GL(3,R)$ shift of $v$
it is possible to transform the model  (4) into the  trivial
one (with flat $g_{ij}$ and constant $b_{ij}, \p$).
Changing the coordinates $(u,v,x_1,x_2)\ra (u,v', r, \theta)$
$$ r \cos{\theta}
=x_1+x_2 \cos{u}\ , \quad  r\sin{\theta}= x_2 \sin{u}\ , \quad v'=v-{1\over
2}x_1x_2 \sin{u}\ ,
$$
one can put (4) into the form
$$I(u,v,r,\theta ) =   {1 \over \pi \a' } \int d^2 z  \bigl(  -2 \del u \bd v'
+
 \del r \bd r  +  r^2 \del \theta   \bd \theta  - 2 r^2 \del u \bd \theta
\bigr) +  {1 \over 4 \pi
}  \int d^2 z  \sqrt \g R^{(2)}
 \p_0 \ . \eq{11} $$
The simple duality with respect to $\theta$ transforms (11) into
$$I'(u,v,r \theta' ) =   {1 \over \pi \a' } \int d^2 z  \bigl(  -2 \del u \bd
v' +
 \del r \bd r  +  r^{-2} \del \theta'   \bd \theta'  -  \ha  \del u \bd \theta'
 - \ha
\bd u \del  \theta'  \bigr)  $$ $$ + \  {1 \over 4 \pi
}  \int d^2 z  \sqrt \g R^{(2)}
 (\p_0 - \ha \ln r ) \ . \eq{11'} $$
If we now shift $v' \ra v + \ha \theta'$ and  perform the duality with respect
to $\theta'$
we get precisely the flat \sm with constant antisymmetric tensor  and dilaton.
 One is thus
 able to relate  the model (4)  to the flat  one
by combining a peculiar coordinate  transformation  with two $O(2,2)$ rotations
 and a
linear  shift  of $v$ (as noted above, the latter can be considered as an
element of $O(3,3)$ rotation).\foot {While typing this paper we have learned
about   a recent preprint
\koki\  in which  a possibility to transform the background of \napwi\ to flat
space by a  pure
$O(2,2)$ duality rotation is mentioned.  We were also informed  that a  duality
rotation of a gauged
version of the model of \napwi\  also leading to a flat space  was
 discussed in  \sfts. }

In conclusion, let us emphasize the important role  played by the antisymmetric
tensor
background in the models (1). Since the duality intertwines the metric and  the
antisymmetric
tensor,  in some cases the antisymmetric tensor background may play
the role  (from a string point of view) of a curved metric background.
We hope that further study of the class of backgrounds (1) may help to
elucidate
the role of duality as a (part of) generalized `coordinate invariance' in
string theory.

\vfill\eject
\listrefs
\end